\documentstyle[11pt,epsfig,amsbsy]{article}

\def\be{\begin{equation}}
\def\ee{\end{equation}}
\def\ba{\begin{eqnarray}}
\def\ea{\end{eqnarray}}

\def\ga{\mathrel{\raise.3ex\hbox{$>$\kern-.75em\lower1ex\hbox{$\sim$}}}}
\def\la{\mathrel{\raise.3ex\hbox{$<$\kern-.75em\lower1ex\hbox{$\sim$}}}}

\newcommand{\fr}[2]{\frac{#1}{#2}}

\newcommand{\oQ}{\omega_{\rm{Q}}}

\newcommand{\Omo}{\Omega_{\rm{m}}^{0}}
\newcommand{\Odeo}{\Omega_{\rm{de}}^{0}}
\newcommand{\Oma}{\Omega_{\rm{m}}(a)}

\newcommand{\Odea}{\Omega_{\rm{de}}(a)}

\newcommand{\ST}{\rm{ST}}

\begin{document}

\baselineskip=16pt
\begin{titlepage}
\begin{center}

\vspace{0.5cm}

\large {\bf Cluster number counts in quintessence models}
\vspace*{5mm} \normalsize

{\bf Seokcheon Lee$^{\,1}$ and Kin-Wang Ng$^{\,2,3}$}

\smallskip
\medskip

$^1${\it School of Physics, Korea Institute for Advanced Study, \\ Heogiro 85, Seoul 130-722, Korea}

$^2${\it Institute of Physics, Academia Sinica, Taipei, Taiwan 11529, R.O.C.} \\

$^3${\it Institute of Astronomy and Astrophysics, \\
 Academia Sinica, Taipei, Taiwan 11529, R.O.C.}

\smallskip
\end{center}

\vskip0.6in

\centerline{\large\bf Abstract}
Even though the abundance and evolution of clusters have been used to study the cosmological parameters including the properties of dark energy owing to their pure dependence on the geometry of the Universe and the power spectrum, it is necessary to pay particular attention to the effects of dark energy on the analysis. Especially, the dark energy dependence on both the volume element and the growth factor has been studied intensively. However, the matter power spectrum is also affected by the dark energy through its normalization. We obtain the explicit dark energy dependent {\it rms} linear mass fluctuation $\sigma_8$ which is consistent with the CMB normalization with less than $2$ \% errors for general constant dark energy equation of state, $\oQ$. Thus, we break the degeneracy between $\sigma_8$ and the matter energy density contrast $\Omo$ to obtain the dark energy dependence. When we use the correct value of the critical density threshold $\delta_{c} = 1.58$ obtained for the spherical collapse model \cite{09090826, 09100126} into the cluster number density $n$ calculation in the Press-Schechter (PS) formalism, $n$ increases compared to the one obtained by using $\delta_{c} = 1.69$ by about $60$, $80$, and $110$ \% at $z = 0$, $0.5$, and $1$, respectively. Thus, PS formalism predicts the cluster number consistent with both simulation and observed data at the high mass region. We also introduce the improved coefficients of Sheth-Tormen (ST) formalism, which is consistent with the recently suggested mass function \cite{10052239}. We found that changing $\oQ$ by $\Delta \oQ = -0.1$ from $\oQ = -1.0$ causes the changing of the comoving numbers of high mass clusters of $M = 10^{16} h^{-1} M_{\odot}$ by about $20$ and $40$ \% at $z = 0$ and $1$, respectively.

\vspace*{2mm}

\end{titlepage}

\section{Introduction}
\setcounter{equation}{0}

The formation of the large scale dark matter (DM) potential wells of clusters is solely determined by gravitational physics irrelevant to the gas dynamical processes, star formation, and feedback. Also clusters are the largest virialized objects in the Universe with their abundance and evolution simply related to the linear matter power spectrum \cite{0002336, 0406331, 0410173, 0506395, 0605575, 11034829}. Thus, the abundance of clusters and their distribution in redshift should be determined purely by the geometry of the Universe and the power spectrum of initial density fluctuations. As a result, the clusters of galaxies provide a useful probe of the fundamental cosmological parameters including the investigation into the dark energy equation of state $\oQ$, because the linear growth factor $D_g$, the cosmological volume element, as well as the primordial scalar amplitude at horizon crossing $\delta_{H}$ depends on $\oQ$ \cite{0002336, 9906174}.

Abundance of rich clusters has been commonly used to constrain the matter power spectrum because it is sensitive to the normalization of the power spectrum on cluster scales \cite{9511007, 9804015}. The amplitude of the matter perturbations is sensitive to the presence of dark energy, through the normalization of the matter power spectrum to the raw large scales probed by the cosmic microwave background (CMB) anisotropies \cite{9607060}.  The normalizations of the matter power spectrum obtained from both methods can be represented by the {\it rms} mass fluctuation on $8 h^{-1}$Mpc scales. However, there exists discrepancy in $\sigma_8$ values resulted from two different normalization methods \cite{0206507}. While $\sigma_8$ is almost constant in the former method, it drops rapidly for larger $\oQ$ because of the increasingly strong integrated Sachs-Wolfe (ISW) effect in the latter.

There have been numerous papers investigating cluster abundances in the quintessential universe \cite{0309485, 0405636, 0504465, 0506043, 0506200, 0605621, 09023226}. Most of them focus on the influences of dark energy on the background evolution and the growth factor. However, we also investigate the dark energy effect on the normalization of the primordial density fluctuation and obtain proper $\sigma_8$ for the cluster abundance. This effect has been ignored in the previous studies and we show that this is not the case. We also use the correct critical threshold density $\delta_{c} = 1.58$ in our analysis. This gives the more accurate and consistent check for the dark energy study through cluster physics.

In this paper, we briefly review the effects of quintessence field on the matter power spectrum given in Ref. \cite{9906174}. It has been commonly assumed that there is no significant effects of quintessence field on the normalization $\sigma_8$ because it clusters gravitationally on large length scales but remains smooth like the cosmological constant on small length scales due to the relativistic dispersion of its fluctuations. However, we directly show the effects of the quintessence field on the $\sigma_8 (M)$ due to the change in $\delta_H$ in the next section. This gives the very important role to investigate dark energy dependence on the cluster physics. We also repeat the effects of quintessence field on both the background evolution and the growth factor which have been studied in the previous works. In Sec. 3, we investigate the resulting changes in the cluster abundance from the changing in $\sigma_8$, volume element, and the growth factor all together. From the improved accuracy on $\sigma_8$, we have the consistent result on the cluster abundance with the proper $\delta_c$ and $\sigma_8$. We use the correct critical value for $\delta_c$ to investigate the validity of both the Press-Schechter (PS) formalism and the Sheth-Tormen (ST) one in the calculation of cluster abundances. The comoving number density of galaxy clusters obtained from the PS formalism with the proper $\sigma_8$ and $\delta_c$ can overcome the known problem obtained PS formalism with the traditionally adopted $\delta_c$ value. We consider the general QCDM models to investigate the dark energy effect on the cluster physics. We conclude in Sec. 4.

\section{Cosmological Consequences of Quintessence Models}
\setcounter{equation}{0}
We briefly review and find the influences of the quintessence field on the cosmological quantities in this section. We consider the cold dark matter with the quintessence field (QCDM) in a spatially flat universe. We limit our analysis on the constant equation of state of the dark energy $\oQ$ which is proper for the late time behaviors of quintessence models \cite{0112221, 0112526}. We are able to apply these solutions to the time-varying $\oQ$ by interpolating between models with the constant $\oQ$ \cite{09072108, 10051770}. We show that dark energy dependence on the linear matter power spectrum appears on both the normalization and the growth factor. The first has been ignored in the previous studies and we obtain the {\it rms} mass fluctuation explicitly.

\subsection{The Power Spectrum and $\sigma_8$}
\setcounter{equation}{0}
The linear perturbation equation for the quintessence field $Q = Q_{0} + \delta Q$ is given by
\be \delta \ddot{Q} + 3 H \delta \dot{Q} + ( k^2 + V_{,QQ} ) \delta Q = -\fr{1}{2} \dot{h} \dot{Q}_{0} \, , \label{deltaQ} \ee where dots mean
the derivatives with respect to the cosmic time, $V_{,QQ} = d^2 V / dQ^2 |_{Q = Q_0}$, and $h$ is the trace of the spatial metric perturbation
\cite{9708069, 9708247}. Thus, the Compton wavelength of the quintessence field above which it clusters gravitationally but remains smooth on smaller
scales is determined from the wavenumber $k_{Q} = 2 \sqrt{ V_{,QQ}}$, where $V_{,QQ}$ is given by
\ba V_{,QQ} &=& \fr{9}{4} \fr{H^2}{c^2} (1-\oQ) \Biggl[ 2 (1+\oQ) - \oQ \Oma \Biggr] + \fr{H^2}{c^2} \fr{1}{1+\oQ} \times \nonumber \\
&& \Biggl[ -\fr{1}{2} \fr{d^2 \oQ}{d \ln a^2} + \fr{1}{4(1+\oQ)} \Bigl(\fr{d\oQ}{d \ln a} \Bigr)^2 + \Bigl(\fr{9}{4} + 3 \oQ + \fr{3 \oQ}{4} \Odea
\Bigr) \fr{d \oQ}{d \ln a} \Biggr] \label{VQQ} \, , \ea where $\oQ$ is the equation of state of $Q$ field, $\Oma$ and $\Odea$ are the energy density
contrasts of the matter and the quintessence, respectively. The linear perturbation of the quintessence field $\delta Q$ grows only on large scales
($k \ll k_{Q}$), and thus the quintessence field clusters and affects the evolution of the matter density perturbation $\delta_{m}$. The Compton wavenumber
is determined by two terms in Eq. (\ref{VQQ}). The second term is due to the time variation of $\oQ$ and disappears when $\oQ$ is a constant.
We investigate the contribution of the second term to $k_{Q}$ when $\oQ$ varies slowly. We adopt the so-called Chevallier-Polarski-Linder (CPL)
parametrization $\oQ = \omega_{0} + \omega_{a} (1-a)$ to study the contribution of the second term compared to the first one \cite{0009008, 0208512}. 
We check that the second term is less than $10$ \% compared to the first one for the slowly varying $\oQ$ cases. Thus, we can safely consider only the first term of the wavenumber for those cases. The effects of the quintessence field on the matter power spectrum and the time evolution of gravitational clustering is parameterized by the shape parameter $\Gamma_{Q} = k_{Q} / h$ in Ref. \cite{9906174} and we adopt this shape parameter.

The linear power spectrum for $\delta_{m}$ in QCDM models is given by \be P(k,a) = A_{Q} k^{n_s} T_{Q}^2(k) \Biggl( \fr{D(a)}{D(a_0)} \Biggr)^2, \label{Pka} \ee where $A_{Q}$ is a normalization, $n_s$ is the spectral index of the primordial adiabatic density perturbations, $T_{Q}(k)$ is the transfer function, $D$ is the linear growth factor, and the present scale factor normalized as $a_{0} = 1$. Even though the transfer function $T_{Q}(k)$ does depend on $\oQ$, its change happens only on large scale $k \leq 0.01$ $\rm{h Mpc^{-1}}$. Thus, the correction on the transfer function does not affect the value of the {\it rms} mass fluctuation. However, the main effect of the quintessence field is on the normalization $A_{Q}$ which can be written as $A_{Q} = 2 \pi^2 \delta_{H}^2 (c/H_{0})^{n_s+3}$, where \cite{9906174}
\ba \delta_{H} &=& 2.05 \times 10^{-5} \alpha_{0}^{-1} (\Omega_{m})^{c_1 + c_2 \ln \Omega_{m}} \exp [ c_{3} (n_s-1) + c_4 (n_s-1)^2 ] \label{deltaH} \, , {\rm with} \\
c_{1} &=& -0.789 |\oQ|^{0.0754-0.211 \ln |\oQ|} \, , c_{2} = -0.118 - 0.0727 \oQ \, , c_3 = -1.037 \, , c_4 = -0.138  \nonumber \, , \\
\alpha &=& (- \oQ)^{s} \,\,\, \rm{with} \,\,\, s = (0.012 - 0.036 \oQ - 0.017 \oQ^{-1}) \Bigl(1 - \Omega_{m}(a) \Bigr) \nonumber \\
&& + (0.098 + 0.029 \oQ - 0.085 \oQ^{-1}) \ln \Omega_{m}(a) \label{alpha} \, . \ea
Note that the notation $\alpha_{0} = \alpha(a_0)$ and $\delta_{H}$ is the amplitude at horizon crossing. The dark energy dependence on $\delta_{H}$ and $D$ smears into the matter power spectrum. And these effects appear on $\sigma_8$. One can use either WMAP 7 year data or Planck satellite mission for the cosmological parameters \cite{10014538, 13035076}. However, we will use WMAP 7 year data in our analysis because the large scale structure analysis have been done based on WMAP 7 year data. Thus, we adopt $\delta_{H}=2.05 \times 10^{-5}$ to
be consistent with WMAP $7$ year data, and the cosmological parameters $\Omo = 0.272$, $h = 0.7$, $n_s = 0.963$, $\Omega_{{\rm b}}^{0} = 0.0456$,
and $\sigma_8 = 0.809$ from WMAP $+$ BAO $+$ $H_{0}$ measurements given in Ref. \cite{10014538}. The {\it rms} linear mass fluctuation at the top-hat smoothing scale $R$ is given by
\be \sigma_{R}^2 (a) \equiv \Biggl< \Biggl| \fr{\delta M}{M(R, a)} \Biggr|^2 \Biggr> = \fr{1}{2 \pi^2} \int_{0}^{\infty} k^2 P(k, a) \Bigl| W(kR)
\Bigr|^2 dk, \label{sigmaR} \ee where the filtering radius $R$ is the Lagrangian radius of a halo of mass $M$ at the present epoch,
$R = (\fr{3M}{4 \pi \rho_{m} (a_0)})^{1/3}$, and $W(kR) = \fr{3}{(kR)^3} \Bigl(\sin [kR] - (kR) \cos [kR] \Bigr)$ is the top-hat window function.
Traditionally, the cluster abundance is used to put a constraint on the dispersion of the density contrast at the scale $8$ ${\rm h^{-1} Mpc}$,
denoted as $\sigma_{8}$ \cite{10043337}. We also denote the mass inside a sphere of radius $R_{8} = 8$ ${\rm h^{-1} Mpc}$ as $M_{8} = 5.95 \times 10^{14} \Omo h^{-1}
\rm{M}_{\odot}$ where $\rm{M}_{\odot}$ means the solar mass and we use the present critical energy density $\rho_{\rm{crit}}^{0} = 2.775 \times 10^{11}
{\rm M}_{\odot} h^2 {\rm Mpc}^{-3}$.

\begin{center}
\begin{figure}
\vspace{1.5cm}
\centerline{\psfig{file=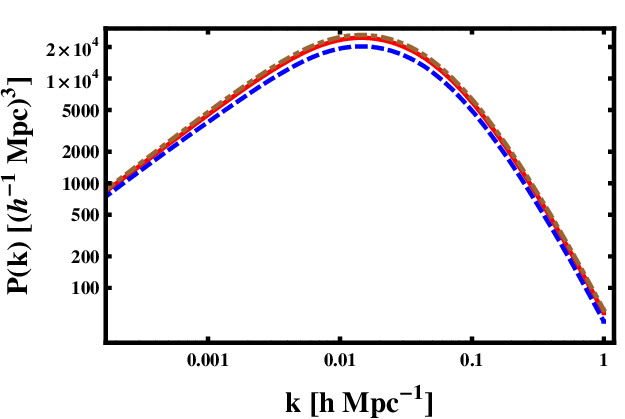, width=6.5cm} \psfig{file=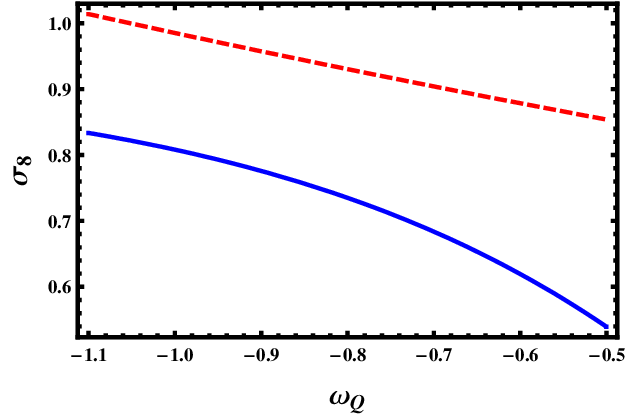, width=6.5cm}}
\vspace{-0.5cm}
\caption{ a) The linear matter power spectrum for different values of $\oQ = -1.1$ (dot-dashed), $-1.0$ (solid), and $-0.8$ (dashed) (from top to bottom). b) $\sigma_8$ values from the COBE normalization (solid) and the Wang and Steinhardt's approximation (dashed) (from bottom to top).} \label{fig1}
\end{figure}
\end{center}

\begin{center}
\begin{figure}
\vspace{1.5cm}
\centerline{\psfig{file=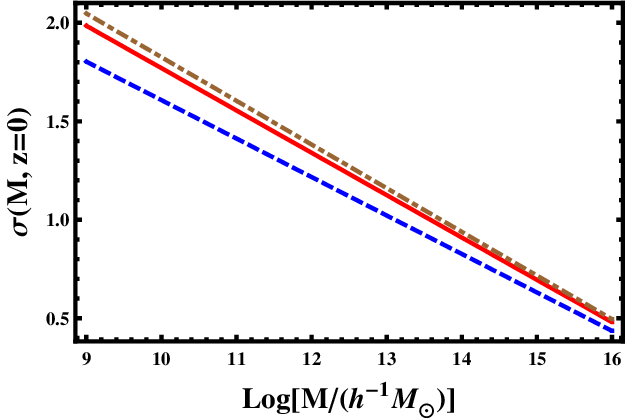, width=8cm}}
\vspace{-0.5cm}
\caption{ $\sigma(M, z=0)$ for different values of $\oQ = -1.1$ (dot-dashed), $-1.0$ (solid), and $-0.8$ (dashed) (from top to bottom).} \label{fig2}
\end{figure}
\end{center}

Now we show the present matter power spectra and $\sigma_8$ values for the different values of $\oQ$ in
Fig. \ref{fig1}. In the left panel of Fig. \ref{fig1}, the dot-dashed, solid, and dashed lines correspond to the matter power spectrum for
$\oQ = -1.1$, $-1.0$, and $-0.8$, respectively. The differences in the power spectra of $\oQ = -1.1$ and $-0.8$ from $-1.0$ are about $7$ and $17$ \%,
respectively. In the right panel of Fig. \ref{fig1}, we also compare the COBE normalized $\sigma_8$ (solid line) with the one from Ref. \cite{9804015} (dashed one). We use Eqs. (\ref{Pka}) - (\ref{sigmaR}) with the cosmological parameters from WMAP 7 to obtain $\sigma_8$ depicted as the solid line in the figure. There exists the discrepancy in the dependence of $\sigma_8$ on $\oQ$ between the two different normalization other than the magnitudes. It is easy to understand this as shown in Ref. \cite{0206507}. For the fixed value of $\Omo$, the dark energy dominates the cosmic expansion earlier as $\oQ$ increases and thus enhancing the
dynamics of the gravitational potential which results in an increasing integrated Sachs-Wolfe (ISW) effect on large scales. This effect is properly
shown in the COBE normalized $\sigma_8$ only. Also $\sigma_8$ obtained from the X-ray cluster population as in Ref. \cite{9804015} suffers from the
uncertainties due to the uncertainties in modeling. Thus, we use the COBE normalized $\sigma_8$ in our cluster abundance calculation.
Even though the COBE normalized $\sigma_8$ is the proper and the accurate one, one needs to express it as a function of $M$ in the cluster abundance
calculation. When $\oQ = -1$, $\sigma_8$ is well known \cite{9604141} and we obtain $\oQ$ dependence on $\sigma_8$ as
$\sigma_8 (\oQ) = (-\oQ)^{0.72 + 0.36 \oQ} \sigma_8 (\oQ = -1)$ with less than $2$ \% errors for $-1.1 \leq \oQ \leq -0.5$. For the cosmological
parameters adopted from the WMAP 7, we obtain $\sigma$ as a function of the cluster mass (M) and the redshift
\be \sigma (M, z) \simeq (-\oQ)^{0.72 + 0.36 \oQ} \Biggl( 3.90 - 0.22 \log \Bigl[\fr{M}{h^{-1} M_{\odot}} \Bigr] \Biggr)
\Biggl( \fr{D_{g}(z)}{D_{g}(z_0)} \Biggr) \, . \label{sigmam} \ee The redshift dependence on $\sigma$ is obtained from the growth factor $D_{g}(z)$. The $\oQ$ dependence comes from the our fitting formula. We improve the $\sigma$ relation with $M$ for the $\Lambda$CDM obtained from the reference \cite{9604141}. We check the consistency of this fitting form by using $M_{8} = 5.95 \times 10^{14} \Omo h^{-1}
\rm{M}_{\odot}$ to obtain $\sigma_8 = 0.81$ for the $\Lambda$CDM. We show $\sigma (M, z=0)$ for different values of $\oQ$ in Fig. \ref{fig2}.
The dot-dashed, solid, and dashed lines correspond to $\oQ = -1.1$, $-1.0$, and $-0.8$, respectively. We check the fitting form shows the proper behavior for both $M$ and $\oQ$. As $M$ is increased, $\sigma$ is decreased.
Also, $\sigma$ is decreased as $\oQ$ is increased.

\subsection{Volume Element and Growth Factor}
\begin{center}
\begin{figure}
\vspace{1.5cm}
\centerline{\psfig{file=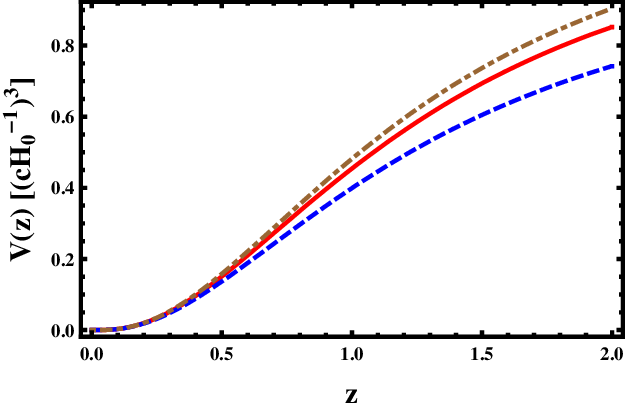, width=6.5cm} \psfig{file=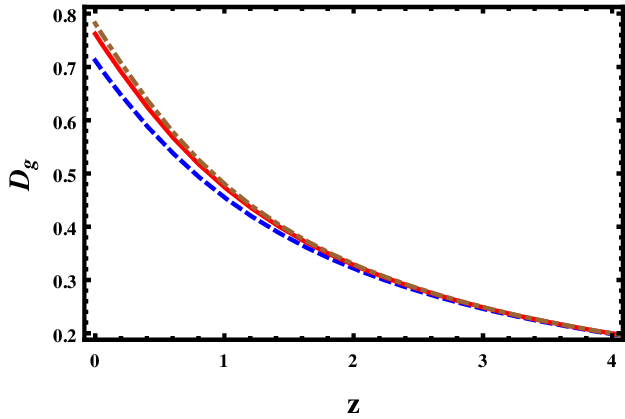, width=6.5cm}}
\vspace{-0.5cm}
\caption{ a) $V(z)$ for different values of $\oQ = -1.1$ (dot-dashed), $-1.0$ (solid), and $-0.8$ (dashed) (from top to bottom). b) $D_{g}(z)$ for the same values of $\oQ$ as in the left panel.} \label{fig3}
\end{figure}
\end{center}

\setcounter{equation}{0}
The different dark energy causes the differences in both the volume element and the growth factor. This have been studied in the previous works and we just repeat it with the correct formula for the growth factor. Friedmann's equation for the QCDM is written as \be H^2(z) = H_{0}^2 \Biggl[ \Omo (1+z)^{3} + \Odeo (1+z)^{3(1+\oQ)} \Biggr]. \label{H2} \ee The cosmic volume per unit redshift is also given by \be V(z) = \int_{0}^{z} 4 \pi d_{A}^2(z') \Biggl| \fr{c dt}{dz} \Biggr|(z') dz' \, , \label{V} \ee where $d_{A} = \fr{c}{1+z} \int_{0}^{z} \fr{dz'}{H(z')}$ is the angular diameter distance between redshifts $0$ and $z$. $V(z)$ is the proper volume of a sphere of radius $z$ around the observer. We show the dependence of $V(z)$ on $\oQ$ in the left panel of Fig. \ref{fig3}. As $\oQ$ increases, $V(z)$ decreases. This is due to the fact that dark energy accelerates the expansion rate and thus the more the negative $\oQ$, the larger the $V$. This is shown in the left panel of Fig. \ref{fig3}. The dot-dashed, solid, and dashed lines correspond to $\oQ = -1.1, -1.0$, and $-0.8$, respectively.
The sub-horizon-scale linear perturbation equation with respect to the scale factor $a$ is well known \cite{Bonnor}, given by \be \fr{d^2 \delta}{da^2} + \Biggl( \fr{d \ln H}{d a} + \fr{3}{a} \Biggr) \fr{d \delta}{d a} - \fr{4 \pi G \rho_{m}}{(aH)^2} \delta = 0 \, . \label{dadelta} \ee The exact analytic growing mode solution $D_g$ of $\delta$ for any value of the constant $\oQ$ is well known \cite{09072108, 09051522, 09061643}: \ba D_g(Y) &=& c_{1} Y^{\fr{3 \oQ -1}{6 \oQ}} F [\fr{1}{2} - \fr{1}{2\oQ}, \fr{1}{2} + \fr{1}{3 \oQ}, \fr{3}{2} - \fr{1}{6 \oQ}, -Y] \nonumber \\ && \, + \, c_{2} F[-\fr{1}{3\oQ}, \fr{1}{2 \oQ}, \fr{1}{2} + \fr{1}{6 \oQ}, -Y] \, , \label{deltask} \ea where $A = \fr{\Omo}{\Odeo}$, $Y = A a^{3 \oQ}$, $F$ is the hypergeometric function, and $c_1$ and $c_2$ are related to each other \ba && \fr{c_1}{c_2} \Bigl(a_{i}, \Omega, \oQ \Bigr) = 2 a_{i}^{\fr{1-3\oQ}{2}} A^{\fr{1}{6\oQ}-\fr{1}{2}} (9\oQ -1) \Biggl( -(1+3\oQ) \times \nonumber \\ && F \Bigl[-\fr{1}{3\oQ},\fr{1}{2\oQ},\fr{1}{2}+\fr{1}{6\oQ},-Y_i \Bigr] \nonumber + 3 Y_i F \Bigl[1-\fr{1}{3\oQ}, 1+\fr{1}{2\oQ}, \fr{3}{2}+\fr{1}{6\oQ}, -Y_i\Bigr] \Biggr) \nonumber \\ && \Bigg/ 3(3\oQ+1)(\oQ-1) \Biggl( Y_i (3\oQ+2) F \Bigl[\fr{3}{2}-\fr{1}{2\oQ},\fr{3}{2}+\fr{1}{3\oQ},\fr{5}{2}-\fr{1}{6\oQ},-Y_i \Bigr] \nonumber \\ && + (1-9\oQ) F \Bigl[-\fr{1}{2\oQ}+\fr{1}{2},\fr{1}{2}+\fr{1}{3\oQ},\fr{3}{2}-\fr{1}{6\oQ},-Y_i \Bigr] \Biggr) \, , \label{c1c2} \ea where $Y_i = a_{i}^{3\oQ} A$ and $a_{i}$ is the initial epoch to satisfy the sub-horizon scale.
In the right panel of Fig. \ref{fig3}, we show the behaviors of the growth factor $D_{g}$ for the different dark energy models ({\it i.e.} for the different values of $\oQ$) when $\Omega_{m}^{0} = 0.272$. The dot-dashed, solid, and dashed lines correspond to $\oQ = -1.1, -1.0$, and $-0.8$, respectively.
As $\oQ$ decreases, $D_{g}$ maintains the linear growth factor proportional to $a$ for a longer time.

\section{Mass Function and Cluster Number}
\setcounter{equation}{0}

\begin{center}
\begin{figure}
\vspace{1.5cm}
\centerline{\psfig{file=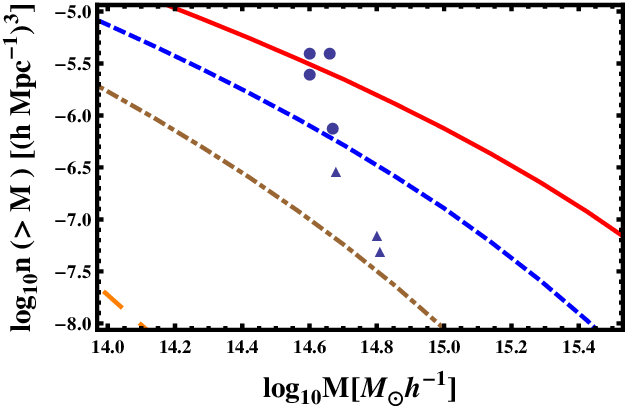, width=6.5cm} \psfig{file=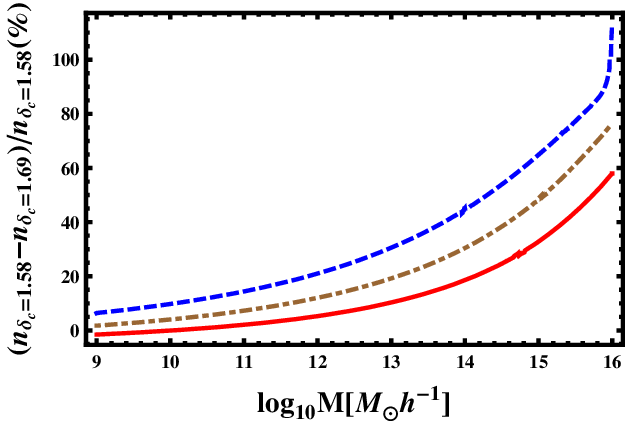, width=6.5cm}}
\vspace{-0.5cm}
\caption{ a) The comoving number density of clusters $n$ of mass greater than $M$ for different values of $z = 0$, $0.5$, $1.0$, and $2.0$ (from top to bottom) when $\oQ = -1$ and $\delta_c = 1.58$. The circular ($z \simeq 0$) and triangular ($0.18 \leq z \leq 0.85$) dots represent the data from Ref. \cite{9612169}. b) Errors of $n$ when we use the correct threshold density contrast $\delta_c = 1.58$ instead of $1.69$ for different values of $z = 0$, $0.5$, and $1.0$ (from bottom to top).} \label{fig4}
\end{figure}
\end{center}
The mass function $f(M)$ from the Press-Schechter (PS) formalism is related to the comoving number density $dn$ of objects in the range $dM$ as $dn = (\rho_{m}^{0} / M) |d \ln \sigma / d M| f(M) dM$ \cite{PS, BCEK}. PS theory which relates the comoving number density of
the virialized objects to their mass is given by
\be dn(M,z) = \sqrt{\fr{2}{\pi}} \fr{\rho_{m}^{0}}{M^2}  \Biggl| \fr{d \ln \sigma}{d \ln M} \Biggr| \fr{\delta_{c}}{\sigma}
\exp \Bigl[ - \fr{\delta_{c}^2}{2 \sigma^2} \Bigr] dM, \label{dn} \ee where the critical density threshold
$\delta_{c} = \fr{\rho_{\rm{linear}}}{\rho_{m}}$ is predicted for a spherical overdensity of radius $R = (3M / 4 \pi \rho_{m})^{1/3}$ and mass $M$
according to the linear theory. However, the PS mass function is known to predict too many low mass clusters and too few high mass clusters,
as well as too few clusters at high $z$ \cite{0504465}. One remark is that the correct value of $\delta_c = 1.58$ was recently obtained independent of
the value of $\oQ$ instead of the well-known value of $1.69$ \cite{09090826, 09100126, SCL}. If we use this correct value of $\delta_c$, then the PS
formalism shows the improved predictions for the number densities of both the low mass clusters and the high mass ones. We show this in Fig. \ref{fig4}.
In the left panel of Fig. \ref{fig4}, we show the comoving number density of clusters $n$ of mass great than $M$ for $\oQ = -1$. The solid, dashed,
dot-dashed, and long dashed lines correspond to $z = 0$, $0.5$, $1.0$, and $2.0$, respectively. The circular ($z \simeq 0$) and triangular
($0.18 \leq z \leq 0.85$) dots represent the data from Ref. \cite{9612169}. We use the rather old data for the comparison with the theoretical prediction. This is due to the fact that the full population of clusters remains largely undiscovered even though cluster surveys at millimeter, optical/near-infrared, and X-ray wavelengths have made good progress. We also show the changes of $n$ at high and low masses at different
redshifts in the right panel of Fig. \ref{fig4}. At the low mass $M = 10^9 [h^{-1}M_{\odot}]$, there is about $2$ \% decrease in $n$ at present.
At the high mass $M = 10^{15}\, (10^{16}) [h^{-1}M_{\odot}]$, $n$ increases about $35\, (58)$ \% today. Also at high $z$, $n$ increases about $50\, (75)$ and $65\, (114)$ \% at
$z = 0.5$ and $1.0$, respectively.

\begin{center}
\begin{figure}
\vspace{1.5cm}
\centerline{\psfig{file=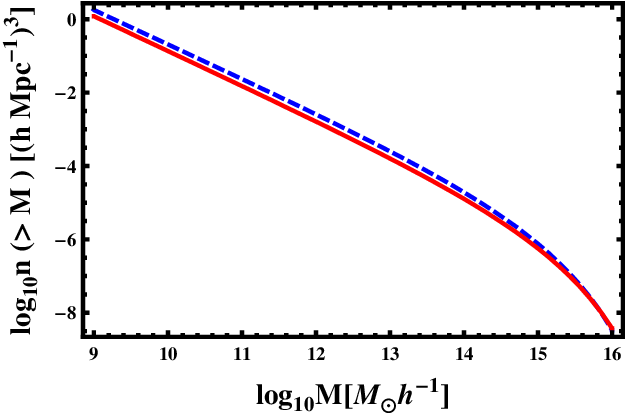, width=8.6cm}}
\vspace{-0.5cm}
\caption{ $n (> M)$ from both PS formalism with correct value of $\delta_c = 1.58$ (dashed line) and ST formalism with $\delta_c = 1.69$ (solid line). } \label{fig5}
\end{figure}
\end{center}

\begin{center}
\begin{figure}
\vspace{1.5cm}
\centerline{\psfig{file=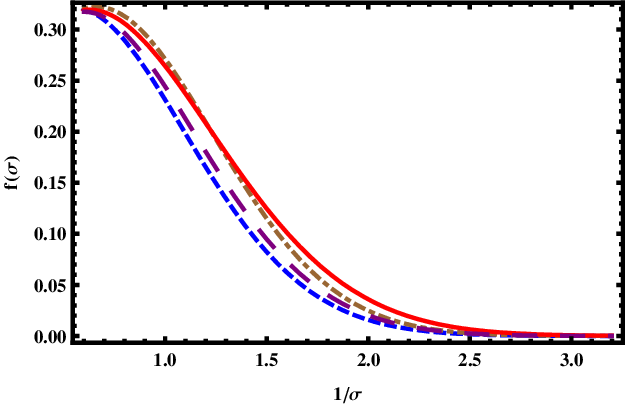, width=8.6cm}}
\vspace{-0.5cm}
\caption{ The different mass functions as a function of $\sigma$: $f_{\rm{ST}}$, $f_{{\rm Manera}}$, $f_{\rm{LN}}$, and $f_{\rm{mod}}$ (from bottom to top). } \label{fig6}
\end{figure}
\end{center}

\begin{center}
\begin{figure}
\vspace{1.5cm}
\centerline{\psfig{file=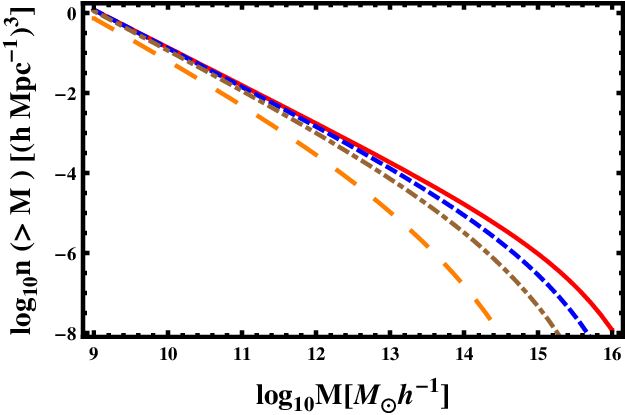, width=6.5cm} \psfig{file=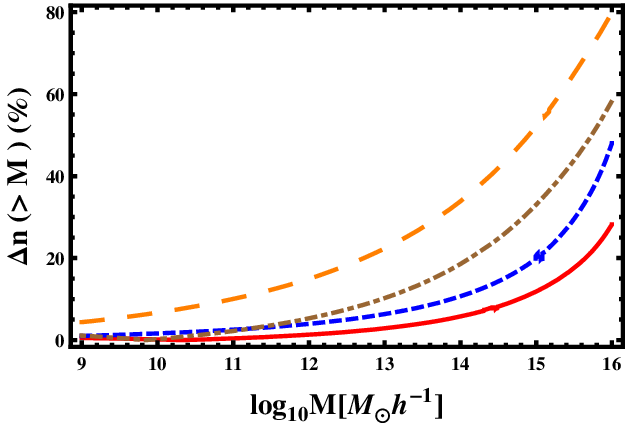, width=6.5cm}}
\vspace{-0.5cm}
\caption{ a) The comoving number density for the mass function given in Eq. (\ref{fLN}) when $\oQ = -1$ at $z = 0$ (solid), $0.5$ (dashed), $1$ (dot-dashed), and $2$ (long dashed) (from top to bottom). b) The relative errors of comoving number density for the two different dark energy models between $\oQ = -1.1$ and $-1.0$ at two different redshifts $z = 0$ (solid) and $z = 1$ (dashed) and for models between $\oQ = -0.8$ and $-1.0$ at $z= 0$ (dot-dashed) and $1$ (long dashed). } \label{fig7}
\end{figure}
\end{center}

Using the correct value of $\delta_c$, we can predict more massive clusters and more clusters at high $z$. However, we may still have too many low mass clusters even by using correct value of $\delta_c$ from the PS formalism.
This is shown is Fig. \ref{fig5}. If we strictly limit the mass of clusters bigger than $10^{14} M_{\odot}$, then PS formalism might be good enough
for the cluster abundance calculation. Also there might be other mechanisms than only gravity for the low mass cluster formations. In any case, PS
formalism shows the deviation from the simulation at low mass region. Thus, we need to consider another popular numerical fit for the differential
mass function given by Sheth and Tormen (ST) \cite{ST99, ST02},
\be f_{\ST}(\sigma) = A \sqrt{\fr{2b}{\pi}} \exp \Biggl[-\fr{b \delta_c^2}{2 \sigma^2} \Biggr]
\Biggl[1 + \Biggl(\fr{\sigma^2}{b \delta_{c}^2} \Biggr)^p \Biggr] \fr{\delta_{c}}{\sigma} \label{fst} \, , \ee
where $A = 0.3222$, $b = 0.75$, and $p = 0.3$ are three parameters tuned to fit with numerical simulations. $A$ is fixed by the normalization that
all dark matter particles reside in halos. However, ST mass function deviates from the simulation results by as much as $40$ \% at the high mass
end \cite{10052239}. Thus, a new fitting function for $f(\sigma)$ is given in Ref. \cite{10052239} by adding one extra parameter into ST,
\be f_{\rm{mod}} (\sigma, z) = \tilde{A} \sqrt{\fr{2}{\pi}} \exp \Biggl[- \fr{\tilde{b} \delta_{c}^2}{2 \sigma^2} \Biggr]
\Biggl[ 1 + \Biggl( \fr{\sigma^2}{\tilde{b} \delta_{c}^2} \Biggr)^{\tilde{p}} \Biggr]
\Biggl( \fr{\delta_c \sqrt{\tilde{b}}}{\sigma} \Biggr)^{\tilde{q}} \label{fmod} \, , \ee
where $\tilde{A} = 0.333 a^{0.11}$, $\tilde{b} = 0.788 a^{0.01}$, $\tilde{p} = 0.807$, and $\tilde{q} = 1.795$. There is an another mass
function $f_{{\rm Manera}}$ which is similar to the original ST formalism \cite{09061314}.
However, the ST mass function is known to be deviated from the simulation at the high mass end \cite{10052239}. This problem can be cured when we use
the correct value of $\delta_c = 1.58$ instead of the traditional value $1.69$. Thus, we introduce a simple but quite similar to the results in Ref. \cite{10052239} by using the original ST formalism:
\be f_{\rm{LN}}(\sigma,z) = 0.32 \sqrt{\fr{2 (0.67)}{\pi}} \exp \Biggl[-\fr{0.67 \delta_c^2}{2 \sigma^2} \Biggr]
\Biggl[1 + \Biggl(\fr{\sigma^2}{0.67 \delta_{c}^2} \Biggr)^{0.32} \Biggr] \fr{\delta_{c}}{\sigma} \label{fLN} \, . \ee The comparison between
the different mass functions $f(\sigma)$s is given in Fig. \ref{fig6}. The dashed, long dashed, solid, and dot-dashed lines correspond to
$f_{\rm{ST}}$, $f_{{\rm Manera}}$, $f_{\rm{LN}}$, and $f_{{\rm mod}}$, respectively. As we can see in this figure, we can produce the enough high mass
clusters from the simple ST formalism by using the correct $\delta_{c}$.
We show the comoving number density of the clusters for $\oQ = -1.0$ at different $z$ in the left panel of Fig. \ref{fig7}. The solid, dashed, dot-dashed, and long dashed lines correspond to $z = 0$, $0.5$, $1.0$, and $2.0$, respectively.
In the right panel of Fig. \ref{fig7}, we show the relative errors of the comoving number densities for two different models at two different
redshifts. For example, the solid line represents the relative errors of $n$ between $\oQ = -1.1$ and $-1.0$ at z = 0,
$| n(>M)_{\oQ = -1.1, z=0} - n(>M)_{\oQ = -1.0, z=0}| / n(>M)_{\oQ = -1.0, z=0} \times 100$ (\%). At low mass $M = 10^9 h^{-1} M_{\odot}$, the
differences of $n$ between two different models ($\oQ = -1.1$ and $-1.0$) are only $0.5$ and $1$ \% at $z = 0$ and $1$, respectively. At this mass,
the differences of $n$ between $\oQ = -0.8$ and $-1.0$ are also only $1$ and $4$ \% at $z = 0$ and $1$, respectively. However, the differences of
$n$ between $\oQ = -1.1$ and $-1.0$ at high mass $M = 10^{16} h^{-1} M_{\odot}$ are $28$ and $48$ \% at $z = 0$ and $1$, respectively. For the two
models $\oQ = -0.8$ and $-1.0$, the differences become $58$ and $80$ \% at two different $z = 0$ and $1$, respectively. Thus, one can investigate the dark
energy consistently through the cluster number density.

\section{Conclusions}
\setcounter{equation}{0}
We investigate the dark energy dependence on the linear matter power spectrum through its normalization and the growth factor. From this, we obtain the mass fluctuation $\sigma_8$ which is consistent with the CMB normalization as a function of the equation of state of the dark energy. Even though we limit our analysis for the constant $\oQ$, we can extend our analysis for the slowly time varying $\oQ$ or use the interpolation of the constant $\oQ$ to study the general time varying $\oQ$. Our formulas for the $\sigma_8$ and the growth factor $D$ make it possible to have the consistent investigation on the dark energy through cluster number density.

We use the correct critical density threshold contrast $\delta_c$ to calculate the comoving number density of clusters for both PS and ST formalism. We show that PS formalism with this correct value of $\delta_c$ can predict the consistent cluster number with both the simulation and the observed data at the high mass region. However, the PS formalism with this correct value of $\delta_c$ still predicts too many low mass clusters. This might be due to other mechanisms for the cluster formation at the low mass region in addition to gravity. We can have the better result with the improved mass function $f_{LN}$. Thus, PS formalism with the correct values of $\delta_c$, $\sigma_8$, and the mass function might be good enough to explain the cluster abundance.

We obtain that the dark energy dependence of the comoving number densities. For the high mass $M = 10^{16} h^{-1} M_{\odot}$ the number density differences between $\oQ = -1.1$ and $-1.0$ are about $28$ and $48$ \% at two different redshifts $z = 0$ and $1$, respectively. For the two
models $\oQ = -0.8$ and $-1.0$, the differences become $58$ and $80$ \% at two different $z = 0$ and $1$, respectively. Thus, observation need to be as accurate as this level to probe the property of dark energy from the cluster physics.


\end{document}